# Thermal Protection System Requirements for Future Planetary Entry and Aerocapture Missions

Athul Pradeepkumar Girija [1,**,***]

[1]School of Aeronautics and Astronautics, Purdue University, West Lafayette, IN 47907, USA

## ABSTRACT

Thermal protection systems are a critical component of planetary exploration, enabling probes to enter the atmosphere and perform in-situ measurements. The aero-thermal conditions encountered during entry are destination and vehicle dependent, ranging from relatively benign conditions at Mars and Titan, to extreme conditions at Venus and Jupiter. The thermal protection system is a single-point-of-failure for both entry probe and aerocapture missions, and hence must be qualified using ground tests to ensure mission success. The high density Carbon-Phenolic which was used in the Galileo and the Pioneer Venus missions is no longer available due to the lack of the manufacturing base for its raw materials. To address the need for Venus and outer planet missions, NASA has developed the Heatshield for Extreme Environment Entry Technology (HEEET). The present study uses the Aerocapture Mission Analysis Tool (AMAT) to perform a comparative study of the thermal protection system requirements for various planetary destinations and the applicability of HEEET for future entry and aerocapture missions. The heat rate and stagnation pressure for aerocapture is significantly less compared to probe entry. The large heat loads during aerocapture present a challenge, but HEEET is capable of sustaining large heat loads within a reasonable TPS mass fraction.

***Keywords:*** Thermal Protection System, Planetary entry probe, Aerocapture

---

[****] To whom correspondence should be addressed, E-mail: athulpg007@gmail.com

# I. INTRODUCTION

NASA has successfully flown numerous planetary entry missions to date, the most demanding of which were the Galileo probe which entered Jupiter and the Pioneer Venus probes, due to their extreme thermal protection system (TPS) requirements [1]. Thermal protection systems are a critical component of planetary exploration, enabling probes to enter the atmosphere and perform in-situ measurements. These measurements such as noble gas abundances are of vital importance in our understanding of the origin and evolution of the Solar System [2]. As the entry probe hits the atmosphere at hypersonic speeds, it encounters aerodynamic heating and structural loads for which the vehicle TPS must be designed to withstand. The aero-thermal conditions encountered during entry are destination and vehicle dependent, ranging from relatively benign conditions at Mars and Titan, to extreme conditions at Venus and Jupiter [3]. In addition to entry probes, thermal protections systems are also a critical component for aerocapture vehicles which enter the atmosphere to decelerate. Aerocapture can greatly reduce the propellant mass required, allowing for smaller and less expensive spacecraft for planetary missions [4, 5]. The thermal protection system is a single-point-of-failure for both entry probe and aerocapture missions, and hence must be qualified using ground tests to ensure mission success. The key variables for thermal protection system design are the maximum stagnation point heat rate, the stagnation pressure, and the integrated total heat load [6]. The heat rate and the stagnation pressure drives the TPS qualification, as it needs to be qualified for the combination of the heat rate and the stagnation pressure. The total heat load drives the TPS mass fraction, which must be minimized to allow for a reasonable scientific payload. Unfortunately, qualification of TPS presents a significant challenge for demanding missions such as outer planets as it is often not possible to perform ground tests under full-flight like conditions due to limitations of the test facilities. A combination of tests under various bounding conditions, along with simulations is used to predict the performance of the TPS under flight-like conditions. The high density Carbon-Phenolic (HCP) which was used in the Galileo and the Pioneer Venus missions is no longer available due to the lack of the manufacturing base for its raw materials [7]. To address the TPS need for Venus and outer planet missions, NASA has developed the Heatshield for Extreme Environment Entry Technology (HEEET) which is 40-50% lighter compared to HCP, and has been fully qualified for a set of conditions to be at TRL 6 [8]. HCP required probes to enter steep, as it could handle very high heat rates only for a short duration, but not a large heat load. HEEET enables shallow entry which reduces deceleration loads, and also enables outer planet aerocapture missions with very large heat loads. The present study uses the Aerocapture Mission Analysis Tool (AMAT) to perform a comparative study of the thermal protection system requirements for various planetary destinations and the applicability of HEEET for future entry and aerocapture missions [9, 10].

## II. VENUS

The thick Venusian atmosphere presents a difficult entry environment for rigid aeroshells, characterized by a combination of high heat rate and stagnation pressure [11]. Figure 1 (top) shows the conditions for Pioneer Venus-like entry probe at two entry flight path angles (γ). The shallow entry (γ = -10) subjects the vehicle to about 2000 W/cm$^2$, and 1.5 atm, while the steep entry (γ = -30) results in 5000 W/cm$^2$ and 5 atm. The steeper entry results in a higher heat flux, but a lower total heat load [12]. The total heat load for the shallow and steep entries are 27 kJ/cm$^2$ and 17 kJ/cm$^2$ respectively. While the Pioneer Venus probes with HCP entered quite steep to minimize the heat load, but encountering very high heat rates, future missions with HEEET can enter at shallower angles. The large heat loads for aerocapture with rigid aeroshells make it unsuitable for aerocapture at Venus [13]. However, using drag modulation aerocapture at Venus is very compelling and has been the subject of several recent studies [14, 15]. Figure 1 (bottom) shows the conditions for a low-ballistic coefficient drag modulation aerocapture system at Venus following the design reference mission [16], which results in a peak heat rate under 150 W/cm$^2$, and a stagnation pressure of only 0.01 atm. Figure 1 shows the enormous advantage offered by low-ballistic coefficient systems for Venus entry. The low-ballistic coefficient system and has applications for delivering independent small landers and orbiters to Venus [17], small satellites as part larger missions [18], and atmospheric sample return missions from the Venusian clouds [19].

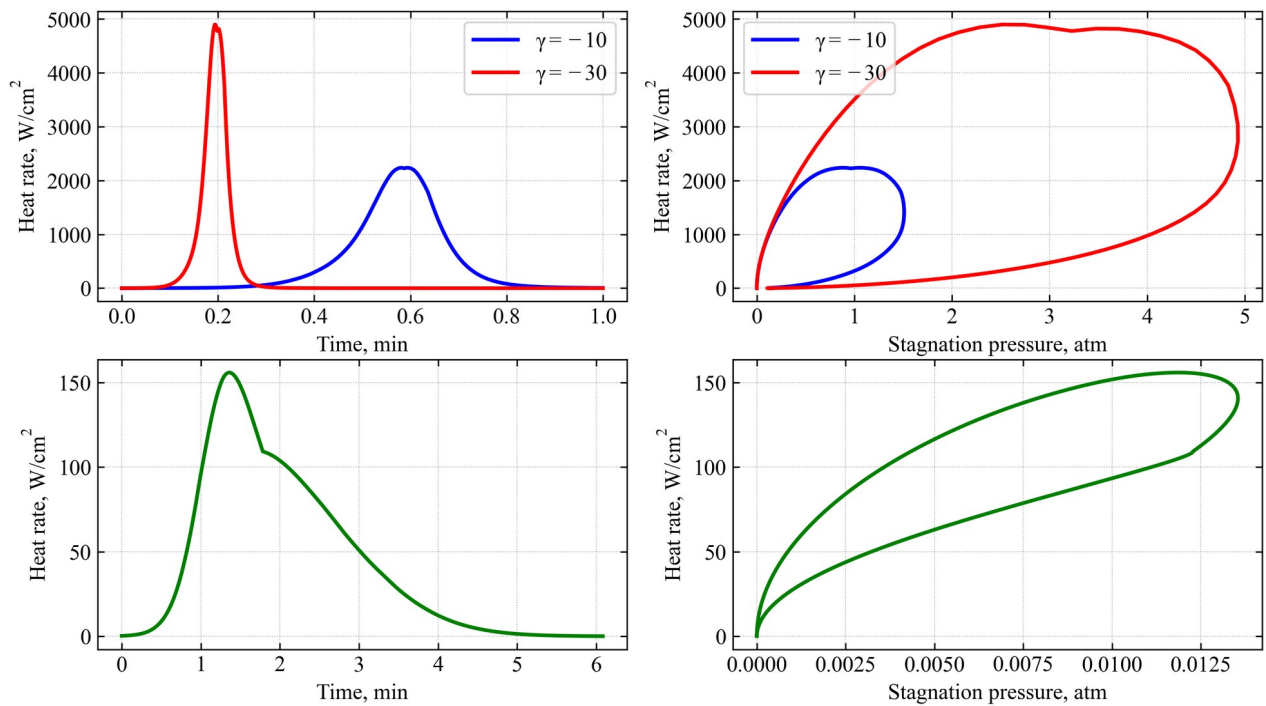

Figure 1. Entry conditions for probe entry (top), and drag modulation aerocapture (bottom) at Venus.

## III. EARTH

The Earth's atmosphere, being not as thick as Venus presents more benign conditions, but still present challenging conditions for high-speed sample entry such as sample return missions. The Stardust sample return capsule for example, entered at -8 deg, 12.5 km/s and using the PICA TPS and encountered about 1000 W/cm$^2$. Future missions such as the Mars Sample Return will enter at higher speeds and will be subject to stringent planetary protection regulations. HEEET will be the likely TPS for such future missions. Figure 2 (top) shows the conditions for Pioneer Venus-like entry probe at two entry flight path angles ($\gamma$). The shallow entry ($\gamma = -8$) subjects the vehicle to about 1000 W/cm$^2$, and 0.5 atm, while the steep entry ($\gamma = -15$) results in 1500 W/cm$^2$ and 1.4 atm. The total heat load for the shallow and steep entries are 34 kJ/cm$^2$ and 23 kJ/cm$^2$ respectively. Aerocapture at Earth has been studied since the 1980s for various technology demonstration missions, but was never flown [20, 21]. However, in recent years there is a renewed interest in a low-cost drag modulation aerocapture technology demonstration mission at Earth [22]. Figure 2 (bottom) shows the conditions for a drag modulation aerocapture at Earth. The peak heat rate and stagnation pressure is about 125 W/cm$^2$, and 0.01 atm., and the total heat load is about 20 kJ/cm$^2$. As with Venus, Figure 2 shows the clear advantage offered by low ballistic coefficient entry systems in terms of the aero-thermal conditions, enabling an order of magnitude lower heat rate compared to rigid aeroshells and 1/100th the stagnation pressure.

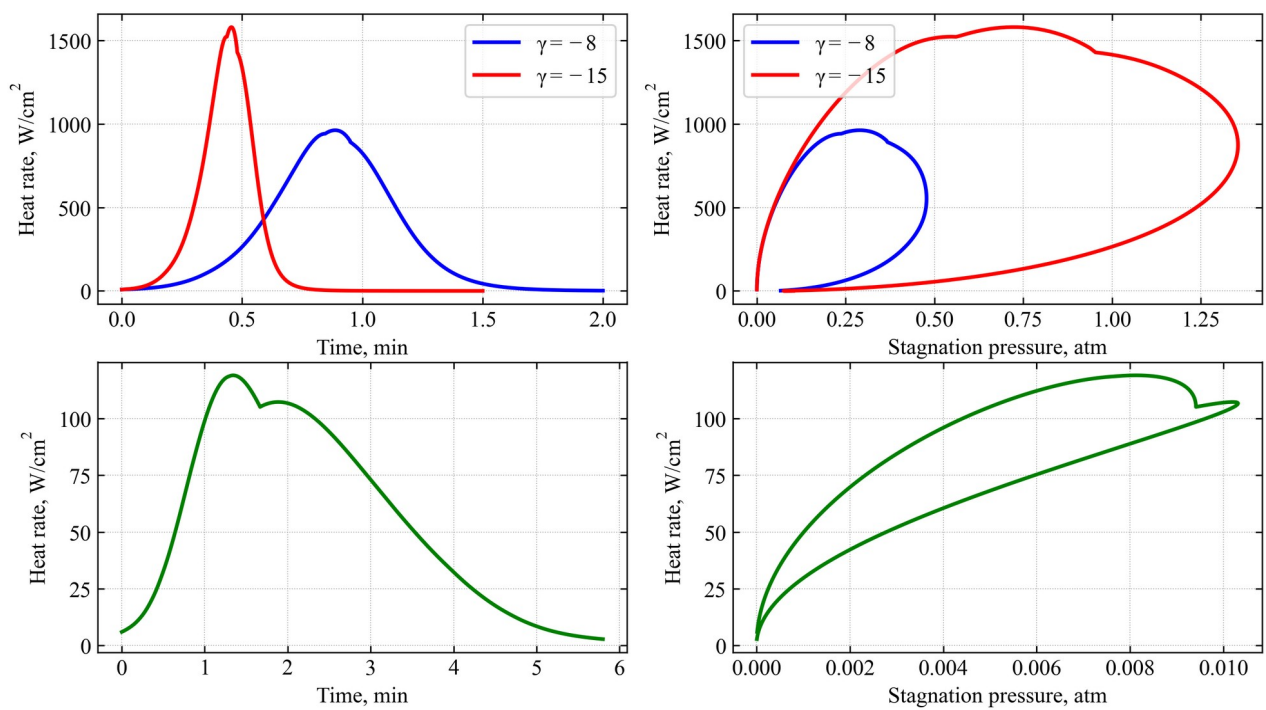

Figure 2. Entry conditions for probe entry (top), and drag modulation aerocapture (bottom) at Earth.

## IV. MARS

With an atmosphere that is far thinner than Venus or Earth, Mars presents the most benign aero-thermal conditions in the inner Solar System though it also presents challenges for slowing down spacecraft before it reaches the surface. Numerous NASA entry vehicles such as the Viking, Pathfinder, and Mars Science Laboratory have landed spacecraft on the Martian surface. Figure 3 (top) shows the conditions for Mars Science Laboratory (MSL) like entry vehicle at two entry flight path angles (γ). The shallow entry (γ = -12) subjects the vehicle to about 40 W/cm$^2$, and 0.08 atm, while the steep entry (γ = -16) results in 50 W/cm$^2$ and 0.17 atm. The total heat load for the shallow and steep entries are 3 kJ/cm$^2$ and 2 kJ/cm$^2$ respectively. These benign conditions enable Mars entry missions to use less dense TPS materials such as PICA, and do not require HEEET. Drag modulation aerocapture has received renewed interest for small satellite orbit insertion at Mars. Mars' low gravity combined and the extended atmosphere makes it more attractive for aerocapture in terms of larger corridor width and lower heating rates compared to Earth or Venus. Figure 3 (bottom) shows the conditions for a drag modulation aerocapture system at Mars. The peak heat rate and stagnation pressure is about 20 W/cm$^2$, and 0.005 atm., and the total heat load is about 5 kJ/cm$^2$. Compared to Figs. 1 and 2, Figure 3 shows the relatively benign entry conditions at Mars. The benign aero-thermal environment and the frequent launch opportunities make Mars an attractive destination for a low-cost aerocapture demonstrator [23].

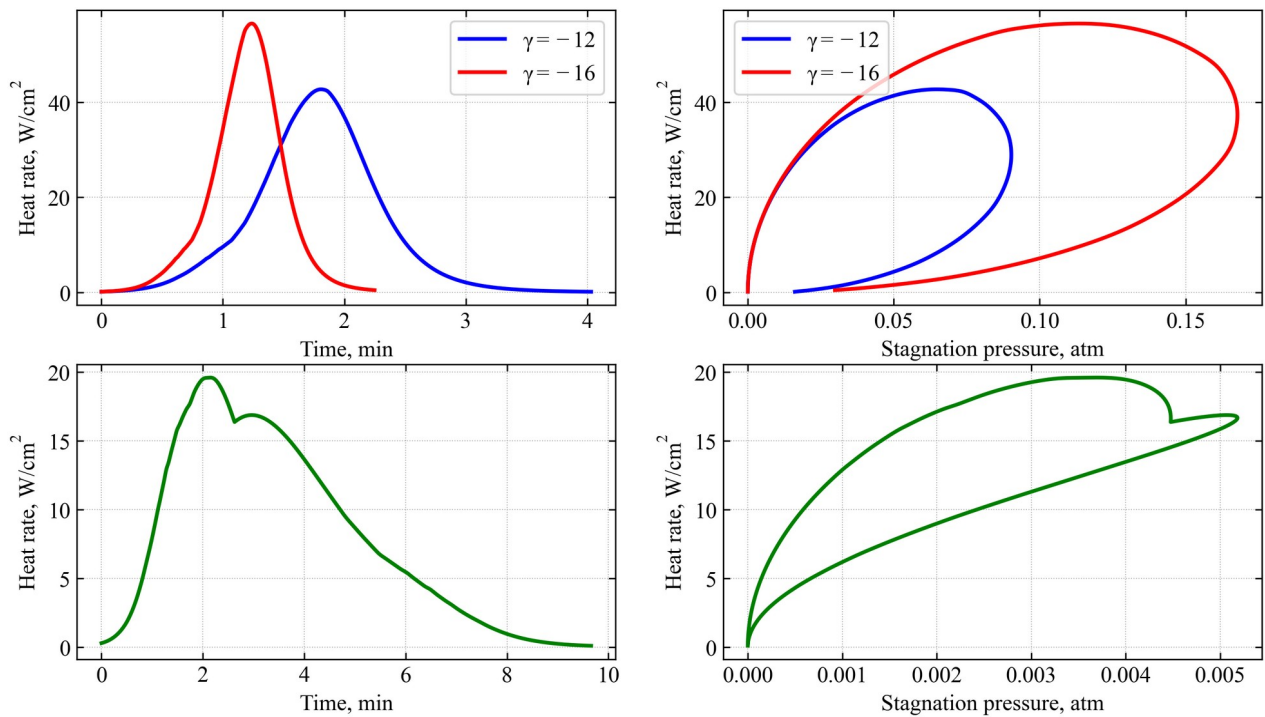

Figure 3. Entry conditions for probe entry (top), and drag modulation aerocapture (bottom) at Mars.

## V. JUPITER AND SATURN

With its enormous gravity well, Jupiter accelerates probes to very high speeds (45 – 55 km/s) as it hits the atmosphere resulting in the most extreme entry conditions anywhere in the Solar System. The Galileo probe entered Jupiter at 47.4 km/s and -8 deg. and remains the most demanding planetary entry ever attempted, encountering heat rates as high as 30,000 W/cm$^2$, about 5 atm and a total heat load of about 200 kJ/cm$^2$. Figure 4 shows the conditions for a Galileo-like like entry vehicle at Jupiter. The conditions are so extreme that it has been compared in the literature to a "ballistic missile warhead flying through a thermonuclear explosion" [7]. Post-flight analysis indicated while that there TPS recession at the stagnation point was less than predicted, at the shoulder which had 5 cm of TPS, there was less than 1 cm of material left suggesting it almost burnt through. This highlights some of the challenges associated with extreme environment entry as the flight-like conditions cannot be achieved in ground facilities and the numerical model predictions may have shortcomings. The conditions at Saturn are less severe than at Jupiter, but is still quite demanding. Figure 5 shows the conditions for a Galileo-like like entry vehicle at Saturn. The shallow entry ($\gamma$ = -12) subjects the vehicle to about 3000 W/cm$^2$, and 1.6 atm, while the steep entry ($\gamma$ = -16) results in 5000 W/cm$^2$ and 2.75 atm. The extreme entry conditions at Jupiter and Saturn make it impractical for aerocapture missions.

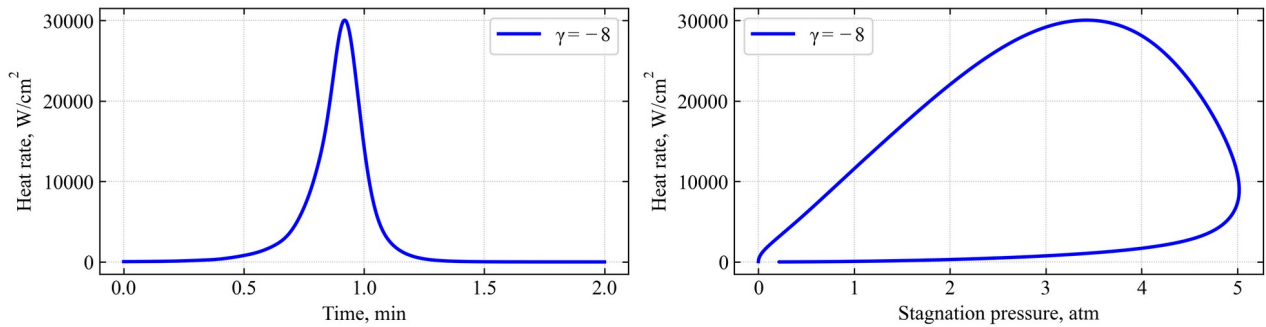

Figure 4. Entry conditions for probe entry at Jupiter.

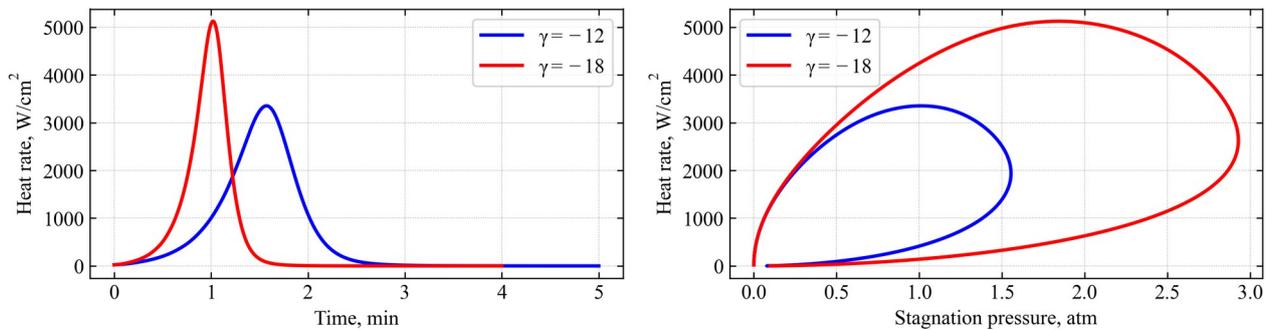

Figure 5. Entry conditions for probe entry at Saturn.

## VI. TITAN

With its low gravity and greatly extended thick atmosphere, Titan is an exception in the outer Solar System and the entry conditions are benign and comparable to Mars. Figure 6 (top) shows the conditions for a Huygens-like like entry vehicle at two entry flight path angles (γ). The shallow entry (γ = -45) subjects the vehicle to about 50 W/cm$^2$, and 0.06 atm, while the steep entry (γ = -65) results in 35 W/cm$^2$ and 0.03 atm. The total heat load for the shallow and steep entries are 2.5 kJ/cm$^2$ and 2 kJ/cm$^2$ respectively. Note the extremely steep entry angles required as the Titan atmosphere is so large, that such steep angles are required for entry. However, the aero-thermal loads are quite small. As with Mars, low density materials such as PICA are more than adequate for Titan entry. The benign entry conditions also make Titan an ideal destination for aerocapture. Figure 6 (bottom) shows the conditions for a drag modulation aerocapture system at Titan. The peak heat rate and stagnation pressure is about 30 W/cm$^2$, and 0.01 atm., and the total heat load is about 5 kJ/cm$^2$ comparable to Mars. Following the Dragonfly mission, there will be a need for an orbiter around Titan to perform global mapping. While propulsive orbit insertion around Titan is extremely challenging due to the large ΔV, aerocapture can potentially enable a Titan orbiter within the New Frontiers cost cap [24]. Titan's benign aero-thermal environment can also be exploited to insert orbiters around Saturn using aero-gravity assist.

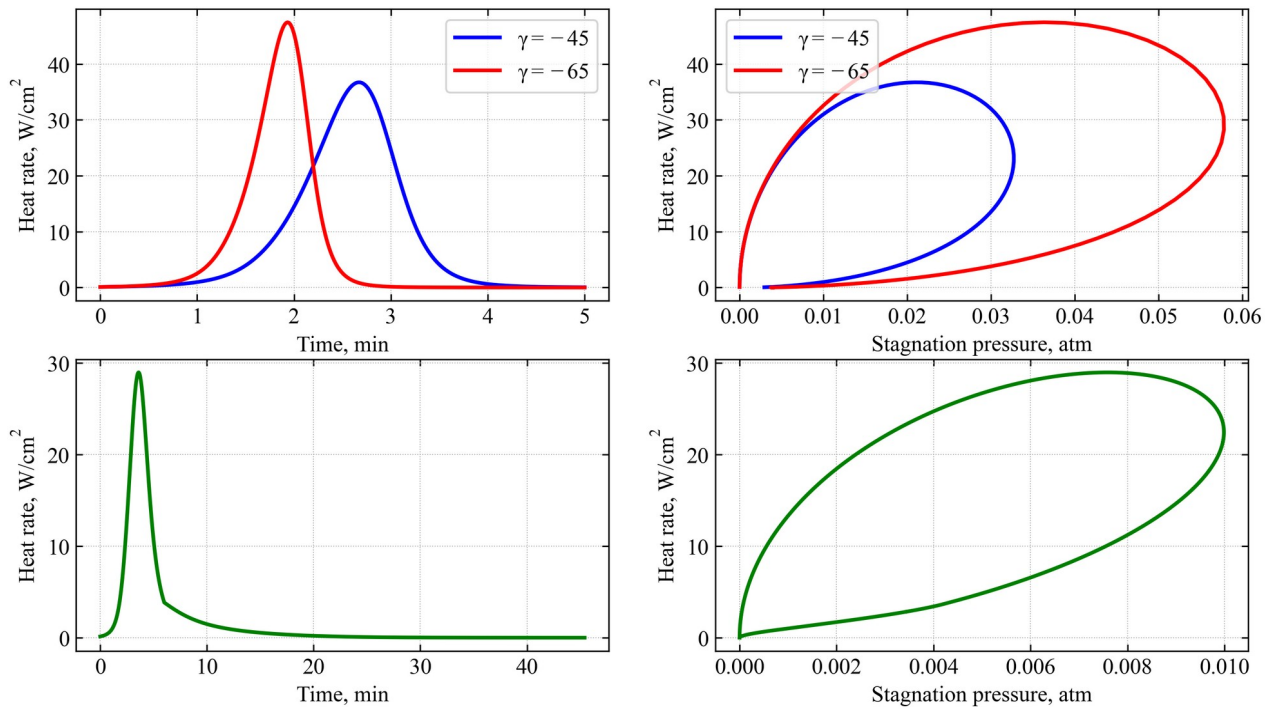

Figure 6.  Entry conditions for probe entry (top), and drag modulation aerocapture (bottom) at Titan.

## VII. URANUS

The 2023-2032 Planetary Science Decadal Survey has recommended the Uranus Orbiter and Probe (UOP) as the highest priority Flagship mission for the next decade. However, the large heliocentric distance (19 AU) and gravity well presents significant challenges. Figure 7 (top) shows the conditions for a probe based on the UOP Decadal Mission study at two entry flight path angles (γ). The shallow entry (γ = -30) subjects the vehicle to about 3000 W/cm$^2$, and 3.5 atm, while the steep entry (γ = -50) results in 3800 W/cm$^2$ and 6 atm. The total heat load for the shallow and steep entries are 54 kJ/cm$^2$ and 40 kJ/cm$^2$ respectively. HEEET is the only material that can withstand these conditions, and has been shown to be 40% more mass efficient compared to HCP. For programmatic reasons, baseline Uranus mission architectures do not consider aerocapture [25, 26]. However, aerocapture can offer significant mission design benefits for Uranus missions [27, 28, 29]. Figure 8 (bottom) shows the conditions for aerocapture at Uranus using an MSL-derived lift modulation vehicle (L/D = 0.24) entering at 29 km/s. The heat rate and stagnation pressure are in the range of 1400 – 2000 W/cm$^2$ and 0.06 – 0.125 atm, which are less demanding than that for an entry probe. However, for shallow entry for aerocapture results in very high heat loads in the range of 200 – 300 kJ/cm$^2$. Such large heat loads are impractical with HCP, as the TPS mass fraction would be too high (> 50%). However, with HEEET it is possible to achieve reasonable TPS mass fractions (< 25%) even under these large heat loads [30].

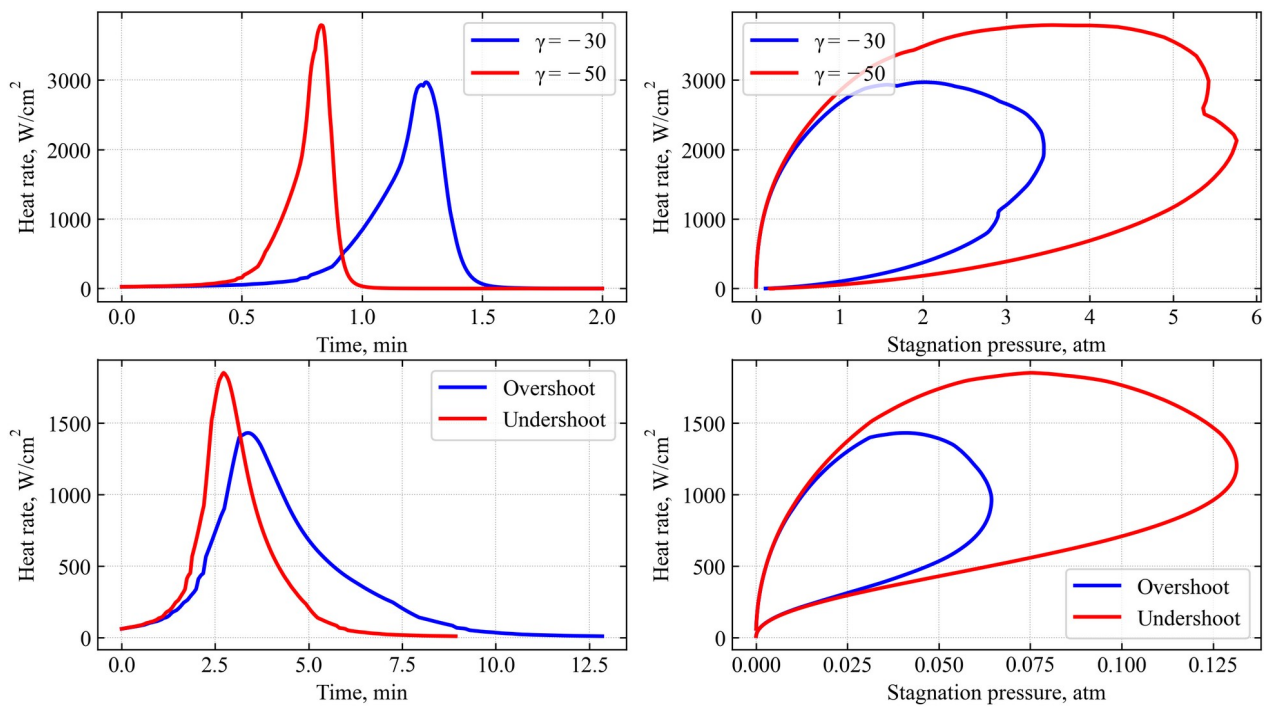

Figure 7. Entry conditions for probe entry (top), and lift modulation aerocapture (bottom) at Uranus.

## VIII. NEPTUNE

At 30 AU, Neptune presents an even greater mission design challenge than Uranus. Its larger heliocentric distance and the associated challenges appear to be the main reason Uranus is prioritized over Neptune for the next Flagship mission. While the two ice giants planets share several similarities, they are also quite different. The most notable is Neptune's moon Triton which may be a captured Kuiper belt object. During Voyager 2's flyby it discovered evidence of geysers, indicating Triton may be an active Ocean World today. Like Uranus, a future Neptune mission will likely include a probe. Figure 8 (top) shows the conditions for a Neptune probe. The shallow entry ($\gamma$ = -30) subjects the vehicle to about 4000 W/cm$^2$, and 6 atm, while the steep entry ($\gamma$ = -40) results in 5000 W/cm$^2$ and 8.5 atm. The total heat load for the shallow and steep entries are 60 kJ/cm$^2$ and 42 kJ/cm$^2$ respectively. As with Uranus, aerocapture offers enormous benefits for future Neptune missions. Early studies in the 2000s had proposed a mid-L/D vehicle to accommodate the large navigation and atmospheric uncertainties at Neptune [31]. However, more recently attention has focused on the use of low-L/D vehicles which have extensive flight heritage and are programmatically more viable than mid-L/D aeroshells [32, 33]. Figure 9 (bottom) shows the conditions for aerocapture at Neptune using an MSL-derived lift modulation vehicle (L/D = 0.24) entering at 30 km/s. The heat rate and stagnation pressure are in the range of 1800 – 2400 W/cm$^2$ and 0.06 – 0.125 atm. The total heat load is in the range of 250 – 320 kJ/cm$^2$.

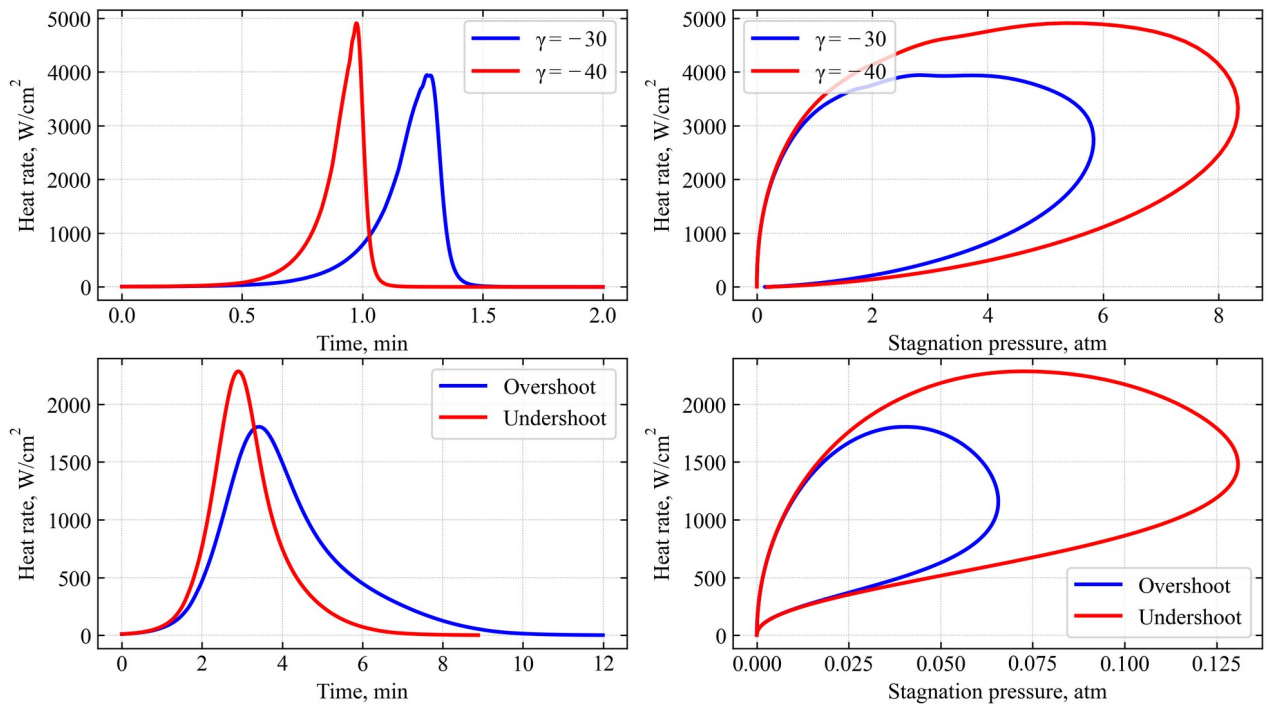

Figure 8. Entry conditions for probe entry (top), and lift modulation aerocapture (bottom) at Neptune.

## IX. COMPARATIVE STUDY OF TPS REQUIREMENTS

Figure 9 compares the TPS requirements for all the entry probes considered in the study. Figure 10 compares a subset of the cases and shows the tested limit of HEEET (3600 W/cm$^2$, 5 atm.) for indicating that HEEET is fully qualified for several missions to Venus, Uranus, and Neptune in the near future. HEEET is potentially capable of even more demanding conditions such as steep Neptune entry, but has not yet been fully qualified for such missions.

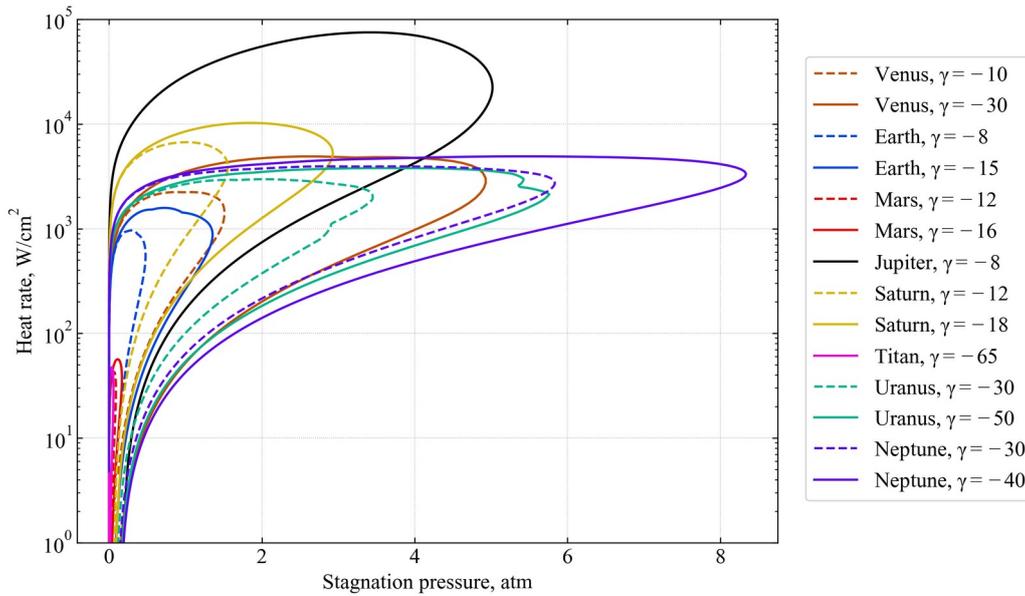

Figure 9. Comparison of the TPS requirements across the Solar System.

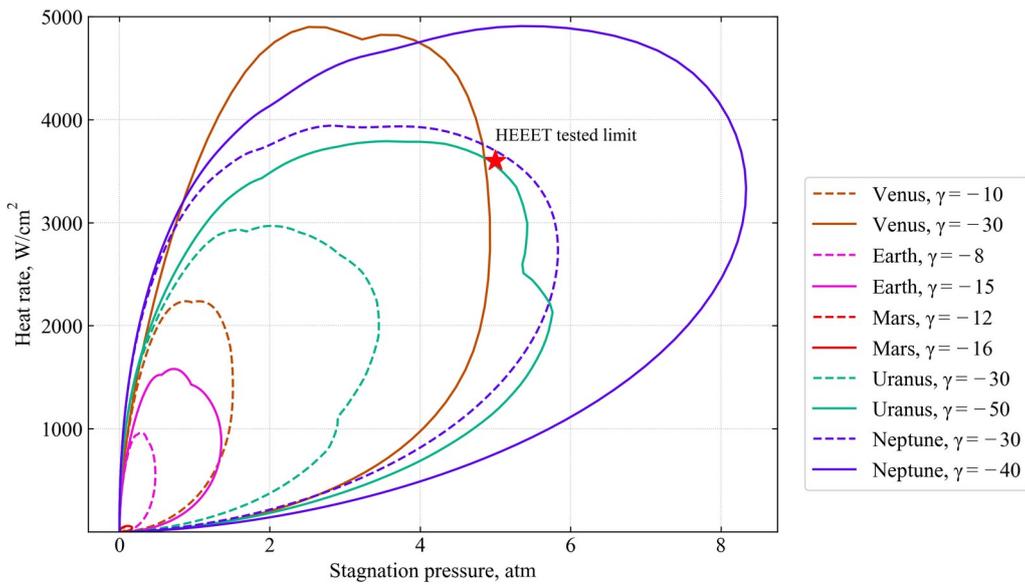

Figure 10. Comparison of the TPS requirements for Venus, Uranus, and Neptune.

Figure 11 compares the TPS requirements for drag modulation aerocapture at Venus, Earth, Mars, and Titan. Figure 12 compares the TPS requirements for lift modulation aerocapture at Uranus and Neptune.

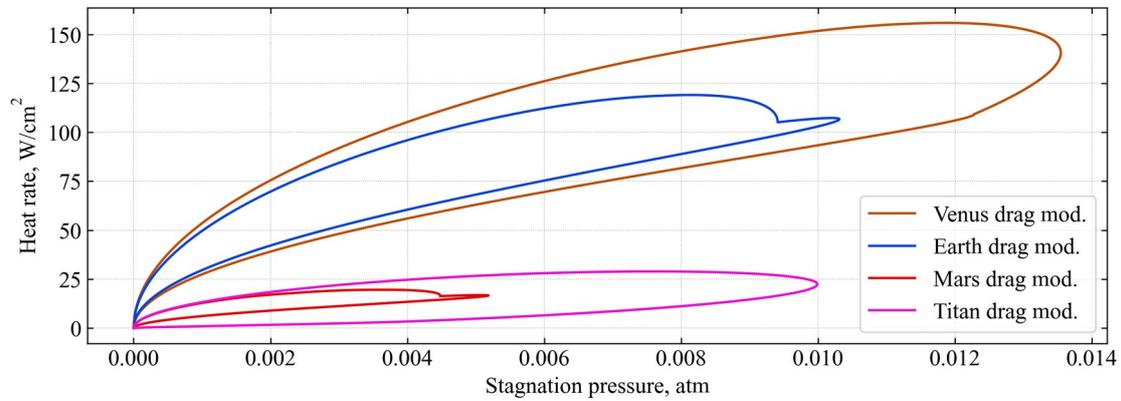

Figure 11. TPS requirements for drag modulation aerocapture.

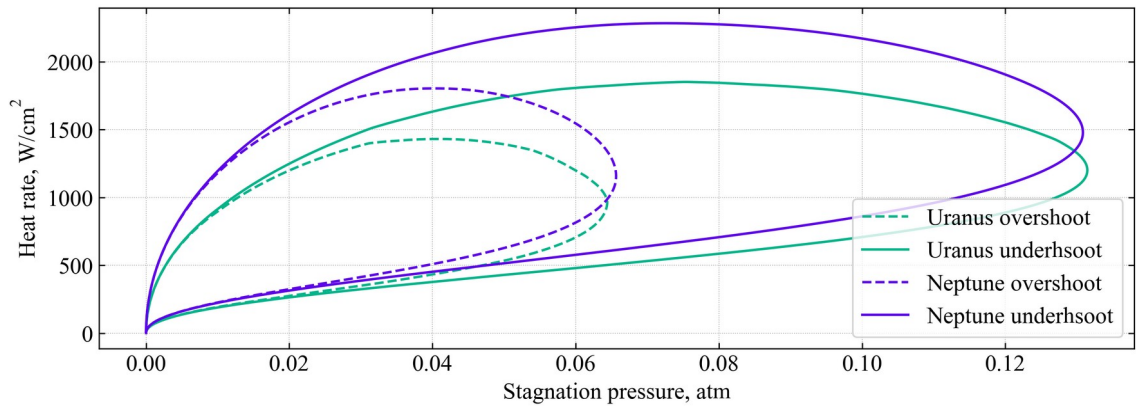

Figure 12. TPS requirements for lift modulation aerocapture at Uranus and Neptune.

## X. CONCLUSIONS

Thermal protection systems are a critical component of planetary exploration, enabling probes to enter the atmosphere and perform in-situ measurements. The present study performed an assessment of the TPS requirements for various planetary destinations and assessed the applicability of the HEEET for future entry and aerocapture missions. HEEET has been tested at 3600 W/cm$^2$, 5 atm. and is qualified to be at TRL 6 for these conditions. The study showed that HEEET enables several future missions such as probe entry at Venus, Uranus, and Neptune which fall within the tested limit. For aerocapture at Uranus and Neptune, the combined heat rate and stagnation pressure is much lower than for probe entry. This makes HEEET which is already qualified for the more demanding probe entry conditions also a viable TPS for aerocapture. The large heat loads encountered during ice giant aerocapture present a challenge, but unlike HCP, HEEET is capable of sustaining large heat loads within a reasonable TPS mass fraction.

## DATA AVAILABILITY

The results presented in the paper can be reproduced using the open-source Aerocapture Mission Analysis Tool (AMAT) v2.2.22. The data and code used to make the study results will be made available by the author upon request.